\documentclass[journal]{IEEEtran}
\usepackage[T1]{fontenc}% optional T1 font encoding
\usepackage{amssymb,amsmath,epsfig,tabularx,delarray,enumerate,subfigure,subfig,graphicx,array,cite,color}
\usepackage {algorithm, algpseudocode}
\usepackage{caption}
\usepackage[cmintegrals]{newtxmath}

\begin{document}
%
% paper title
% Titles are generally capitalized except for words such as a, an, and, as,
% at, but, by, for, in, nor, of, on, or, the, to and up, which are usually
% not capitalized unless they are the first or last word of the title.
% Linebreaks \\ can be used within to get better formatting as desired.
% Do not put math or special symbols in the title.
\title{Nanorobots-assisted Detection of Multifocal Cancer: A Multimodal Optimization Perspective}
%
%
% author names and IEEE memberships
% note positions of commas and nonbreaking spaces ( ~ ) LaTeX will not break
% a structure at a ~ so this keeps an author's name from being broken across
% two lines.
% use \thanks{} to gain access to the first footnote area
% a separate \thanks must be used for each paragraph as LaTeX2e's \thanks
% was not built to handle multiple paragraphs
%

\author{Shaolong~Shi,~\IEEEmembership{Student~Member,~IEEE,}
        Yifan~Chen,~\IEEEmembership{Senior~Member,~IEEE,}
        and~Xin~Yao,~\IEEEmembership{Fellow,~IEEE}% <-this % stops a space
\thanks{S. Shi is with the Harbin Institute of Technology, Harbin 150001, China,
and also with the Department of Electrical and Electronic Engineering,
Southern University of Science and Technology, Shenzhen 518055,
China (e-mail: shishaolong@hit.edu.cn).}% <-this % stops a space
\thanks{Y. Chen is with the Faculty of Science and Engineering and the Faculty
of Computing and Mathematical Sciences, University of Waikato, Hamilton
3240, New Zealand, and also with the Department of Electrical and
Electronic Engineering, Southern University of Science and Technology,
Shenzhen 518055, China (e-mail: yifan.chen@waikato.ac.nz).}% <-this % stops a space
\thanks{X. Yao is with the Department of Computer Science and Engineering,
Southern University of Science and Technology, Shenzhen 518055,
China, and also with the School of Computer Science, University of
Birmingham, Birmingham B15 2TT, U.K. (e-mail: yaox@sustc.edu.cn).}}

% The paper headers
\markboth{Draft paper for IEEE Transactions on Cybernetics}%
{Shell \MakeLowercase{\textit{et al.}}: Bare Demo of IEEEtran.cls for IEEE Journals}

% make the title area
\maketitle

% As a general rule, do not put math, special symbols or citations
% in the abstract or keywords.
\begin{abstract}
We propose a new framework of computing-inspired multifocal cancer detection procedure (MCDP). Under the rubric of MCDP, the tumor foci to be detected are regarded as solutions of the objective function, the tissue region around the cancer areas represents the parameter space, and the nanorobots loaded with contrast medium molecules for cancer detection correspond to the optimization agents. The process that the nanorobots detect tumors by swimming in the high-risk tissue region can be regarded as the process that the agents search for the solutions of an objective function in the parameter space with some constraints. For multimodal optimization (MMO) aiming to locate multiple optimal solutions in a single simulation run, niche technology has been widely used. Specifically, the Niche Genetic Algorithm (NGA) has been shown to be particularly effective in solving MMO. It can be used to identify the global optima of multiple hump functions in a running, keep effectively the diversity of the population, and avoid premature of the genetic algorithm. Learning from the optimization procedure of NGA, we propose the NGA-inspired MCDP in order to locate the tumor targets efficiently while taking into account realistic \emph{in vivo} propagation and controlling of nanorobots, which is different from the use scenario of the standard NGA. To improve the performance of the MCDP, we also modify the crossover operator of the original NGA from crossing within a population to crossing between two populations. Finally, we present comprehensive numerical examples to demonstrate the effectiveness of the NGA-inspired MCDP when the biological objective function is associated with the blood flow velocity profile caused by tumor-induced angiogenesis.
\end{abstract}

% Note that keywords are not normally used for peerreview papers.

\IEEEpeerreviewmaketitle

\section{Introduction}

\subsection{Background}
Cancer refers to the abnormal and uncontrolled cell growth due to an accumulation of specific genetic and epigenetic defects \cite{schulz2005molecular}. It has been a leading cause of death globally for many years \cite{abubakar2015global}. As many cancers are diagnosed only after they have metastasized throughout the body, effective and accurate methods of early-stage cancer detection and clinical diagnosis are urgently needed \cite{bohunicky2011biosensors}. Early detection of tissue malignancy can increase greatly the chances for successful treatment and it is particularly relevant for cancers of breast, cervix, mouth, larynx, colon and rectum, and skin. For example, the five-year survival rate is 90\% if colorectal cancer is diagnosed while still localized (i.e., confined to the wall of the bowel) but reduces to only 68\% for regional disease (i.e., disease with lymph node involvement) and only 10\% if distant metastases are present \cite{ries2006seer}. The conventional imaging technology of cancer detection such as ultrasound, computed tomography, positron emission tomography, and magnetic resonance imaging (MRI) have limitations in terms of tumor grading and molecular characterization as they are not designed to image a small number of cancer cells. Therefore, the conventional imaging modalities are unpromising in early detection and localization of cancer \cite{frangioni2008new,pakzad2006role,bipat2005colorectal}.\par
  Nanoparticles are organic or inorganic, metallic, magnetic or even polymeric particles with a diameter of less than 100 nm \cite{perfezou2012cancer}. Nanoparticles exhibit several unique properties that can be applied to develop chemical and biological sensors possessing desirable features like enhanced sensitivity and lower detection limits\cite{khanna2008nanoparticle-based}. They can cross barriers such as blood-brain barrier or gastrointestinal barrier, which is a major advantage for the detection and visualization of tumor cells at very early stages, ideally at the level of a single cell or multiple cells \cite{perfezou2012cancer}. For example, Liu \emph{et al.} developed a nanosensor that detected Carcinoembryonic Antigen as a model protein, which showed potential applications for early diagnosis of diseases \cite{liu2010highly}. Several studies have also been carried out on early detection of tumor cells \emph{via} the attachment of proteins expressed in very high quantity on cancer cells \cite{zhang2004three,lyman2005american}. These novel nanoscale platforms possess excellent physiochemical properties and offer prolonged circulation times and improved absorption rates. However, these nanoparticles are circulated systemically and thus exhibit poor targeting efficiency for precise localization of the lesion \cite{kingsley2006nanotechnology:,bao2013multifunctional}. Enhancing the diagnostic efficacy of nanoparticles necessitates the use of guidance technologies, which demand effective controlling and tracking of nanoparticles. This, in turn, calls for the integration of sensing and actuation mechanisms at the nanoscale.\par
  Recent decades have witnessed an enormous progress in manipulation of nanorobots in the \emph{in vivo} environment for externally controllable early cancer detection \cite{okaie2014modeling}. Due to their small size, however, individual nanorobots are limited in their operational ranges and sensing capabilities. Nanorobots must work together to cover large areas and perform complex functionalities, which can be achieved through swarm intelligence \cite{okaie2014modeling,garnier2007the,akyildiz2008nanonetworks:}. Furthermore, so far, very few groups have succeeded in developing a fully autonomous (i.e. without the need for an external power source for propulsion) nanorobot capable of effective thrust in practical applications. To overcome these constraints, bio-inspired, biocompatible, and biodegradable nanorobots such as flagellated magnetotactic bacteria (MTB) with nanometer-sized magnetosomes have been proposed \cite{martel2009mri,Chen2016GreenTN}. A swarm of MTB can be guided and tracked by an external monitoring device such as a customized MRI system to deliver drug-containing nanoliposomes to the target areas, which can significantly improve the therapeutic index of various nanocarriers in tumor regions as demonstrated in \cite{okaie2014modeling}. However, to the best of our knowledge, there has not been any experimental work in the existing literature on applying nanorobots for effective early cancer detection. This may be due to the fact that the destination is known \emph{a priori} in targeted drug delivery, which provides a clear direction for generating the guiding field. On the other hand, the knowledge of the location (or even presence) of tissue malignancy is not available in early cancer detection, which significantly increases the complexity of system design.\par
  \par 
  \begin{figure} [!htp]
\begin{center}
  \epsfig{file=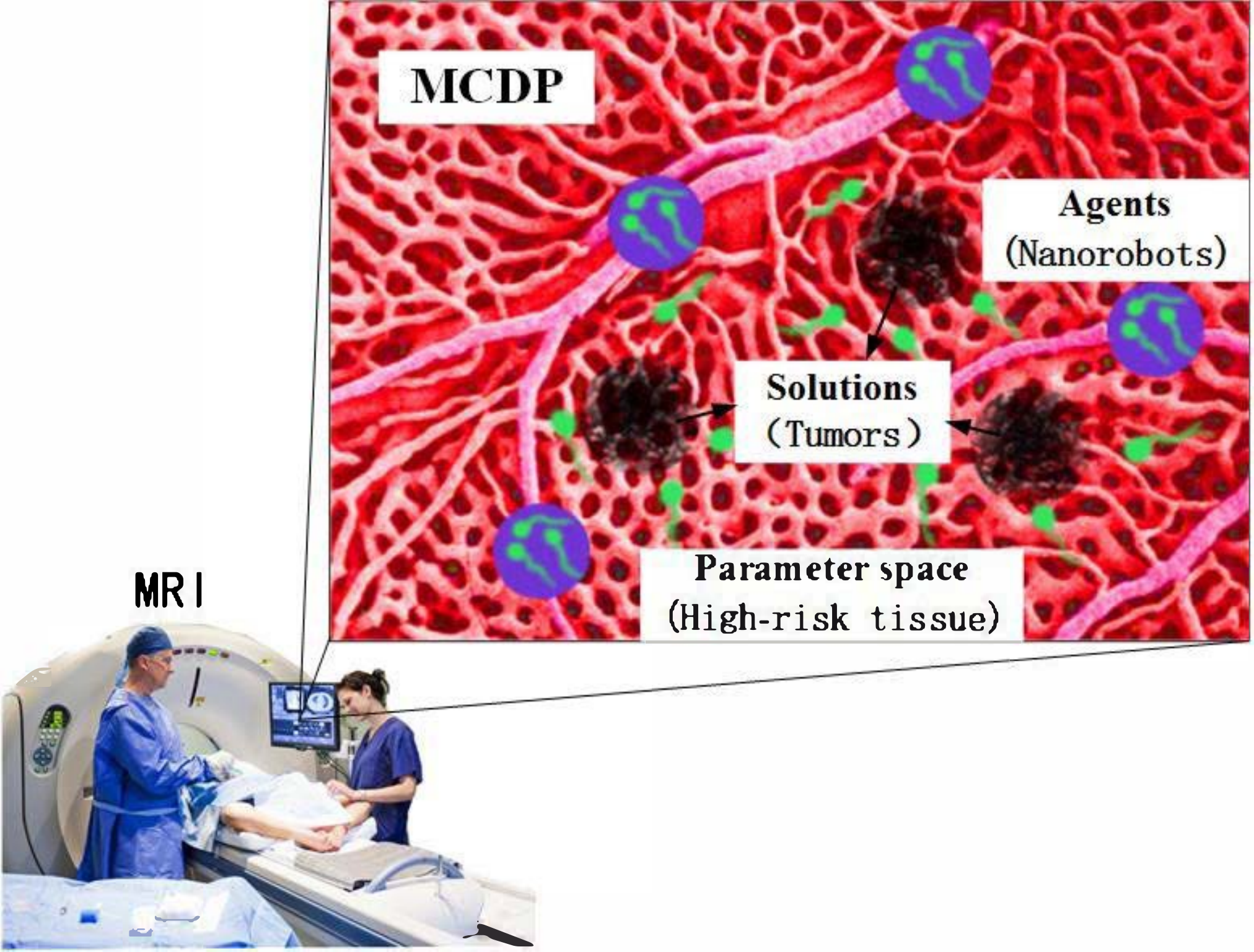,width=1.0\linewidth}
\end{center}
      \caption{Pictorial illustration of the MCDP model}
      \label{fig:Fig.1}
  \end{figure}
  Recently, we have proposed a novel framework of computing-inspired bio-detection, also called ``Touchable Computing'' (TouchComp) \cite{sanap2011nanorobots}, which attempts to shed some theoretical insight on nanorobots-assisted, ``smart'' tumor sensing. TouchComp extends the idea of swarm intelligence to externally controllable biosensing, and improves the detection efficiency compared to the random search strategy relying on systemic circulation without making use of any intelligence about the biological, chemical or physical phenomenon induced by cancer cells\cite{chen2017touchable,martel2009mri,vartholomeos2011mri}. The central idea of TouchComp is summarized in Fig. 1. Consider the case of a multifocal cancer detection procedure (MCDP). The tumor foci are the regions-of-interest (ROIs) in a high-risk tissue region. The tumors trigger certain \emph{in vivo} phenomena that can be detected by observing the properties of a swarm of nanorobots deployed due to the interaction between the nanorobots and the environment. TouchComp aims to coordinate the movement of all nanorobots through specific schemes of swarm intelligence in order to search for the ROIs effectively. 

\subsection{Main Contributions of the Current Work}
Different from our previous work, which focuses on the detection of a single cancer\cite{chen2017touchable}, we look into the multimodal bio-detection scenario (i.e., MCDP) in our present work. Multifocal tumors are not multiple tumors; they originate from a unique cellular clone and grow multifocally in a single organ (liver, kidney, thyroid, etc.)\cite{berg2000multicentric}. As shown in Fig. 1, the solutions are the tumor foci to be detected; the parameter space representing all possible solutions is the tissue region at a high risk of malignancy; the externally manoeuvrable agents are nanorobots such as MTB. The agents (i.e., nanorobots) locate all the optimal solutions (i.e., tumors) by moving through the parameter space (i.e., high-risk tissue) under the guidance of an exterior steering field (i.e., MRI).\par
Different from the classical mathematical computing that uses ideal non-interacting agents, the landscape for the MCDP that corresponds to a biological, chemical or physical cost function induced by tumors may be altered by agents due to the \emph{natural computing} feature of MCDP, where agents interact with the solution space (i.e., nanorobots undergo reactions in the \emph{in vivo} environment). An external observer can then infer the landscape for cancer detection by monitoring the movement of agents, which is called ``seeing-is-sensing'' in \cite{Chen2016GreenTN}. Provided with this analogy, a wide variety of computational techniques can thus be applied to the design of optimal MCDP. \par

 The genetic algorithm (GA) is a metaheuristic inspired by the process of natural selection, which generates high-quality solutions to optimization and search problems by relying on bio-inspired operators such as mutation, crossover and selection \cite{de2006evolutionary}. Niching methods allow the GA to maintain a population of diverse individuals, which can be used in domains that require the maintenance of multiple solutions \cite{mahfoud1995niching}. In this paper, the niche GA (NGA) is considered and introduced into the MCDP with the assumption that the landscape of the tissue region under surveillance is nonlinear and multimodal. Furthermore, we modify the crossover strategy of the standard NGA by dividing the initial population into two groups and implementing crossover between groups to improve the performance of tumor detection. \par
   This paper is organized as follows. In Section II, we establish the foundation of computing-inspired MCDP by transforming the problem into an optimization procedure. In particular, we employ externally controllable nanorobots moving in the blood vessels to indicate the blood flow velocity that cannot be imaged by conventional imaging techniques because of their low resolutions\cite{frangioni2008new,pakzad2006role,bipat2005colorectal,martel2009mri}.  A multiple hump function with three global minima is employed to represent the distribution of blood flow velocity in the presence of cancerous tissues as described in \cite{komar2009decreased,ellegala2003imaging,de2002microvascular}. In Section III, we propose the NGA-inspired MCDP following the problem setting presented in Section II and modify the normal NGA from the algorithmic perspective. In Section IV, some numerical examples are provided to demonstrate the principles and effectiveness of the proposed framework. Finally, some conclusive remarks are drawn in Section V.

\section{Computing-inspired MCDP}
The framework of computing-inspired bio-detection builds upon the similarity between the evolutionary computational process and the nanorobots-assisted MCDP. An evolutionary system can be seen as a process that, given particular initial conditions, follows a trajectory over time through a complex evolutionary state space to develop problem solvers automatically \cite{de2006evolutionary}. The swarm of nanorobots can be seen as an evolutionary system with the aim to accomplish bio-detection in an efficient and robust manner. More specifically, an aggregation of nanorobots loaded with contrast medium molecules are injected into the high-risk tissue, which corresponds to the initialization step of the evolutionary system. The nanorobots swim in the \emph{in vivo} environment following certain trajectories to search for the tumors and the process is monitored by an external macro-scale device.  To evaluate the chance of finding tumors by nanorobots at specific locations, the concept of  ``fitness function'' in mathematics and computer science is introduced. As the existing experimental results show that the velocity of blood flow in which nanorobots exist usually increases with the distance from a cancer as to be elaborated further in Section II-B, the ``fitness'' of a nanorobot is negatively correlated with the blood flow velocity it experiences. Subsequently, the nanorobots update their locations in the tissue region under surveillance according to their fitness values. When a tumor is detected by a nanorobot, the nanorobot will adhere to the tumor and appears to stop moving (i.e., its velocity observed by the external monitoring system reduces to zero). It is worth noting  that the reduction of blood flow velocity alone does not necessarily mean that there is a tumor; only a complete stop of a nanorobot indicates the presence and location of a tumor. Subsequently, the nanorobots-assisted MCDP can be formulated as a stylized representation of the general problem of agents-aided solution-searching in the parameter space\cite{chen2017touchable}.

\subsection{Problem Formulation}
Consider a general parameter space $\mathcal{P}$ with the following agent-dependent landscape \cite{chen2017touchable}:\par

\begin{equation}
\begin{split}
o\left(\vec{x};A\right)&=o_{ms}\left(\vec{x};A\right)+o_{fit}\left(\vec{x};A\right)\\
&=o_{in}\left(\vec{x}\right)+o_{ex}\left(\vec{x};A\right)+o_{fit}\left(\vec{x};A\right),\vec{x}\in\mathcal{P},
\end{split}
\end{equation}
where $o_{ms}\left(\vec{x};A\right)$ is the externally measurable objective function at location $\vec{x}$ for agent $A$, $o_{in}\left(\vec{x}\right)$ is the intrinsic objective function at $\vec{x}$ independent of the presence or absence of $A$, $o_{ex}\left(\vec{x};A\right)$ is the extrinsic disturbance to the landscape caused by the interaction between $A$ and the parameter space $\mathcal{P}$, and $o_{fit}\left(\vec{x};A\right)$ is the correction factor accounting for the  \emph{activeness} of $A$, due to its degradation in $\mathcal{P}$.
% needed in second column of first page if using \IEEEpubid
%\IEEEpubidadjcol
The computing-inspired model of MCDP can be expressed as follows: given a search space (i.e., high-risk tissue) $\mathcal{P}$, the agents move through the search space to locate the targets (i.e., tumors). It is assumed that the tumor locations  remain unchanged regardless of any variation caused by agents to the landscape. Nanorobots can bind to tumor cell receptors when they detect the tumors and stop moving, indicating termination of the MCDP. Furthermore, we use the velocity of blood flow as the objective function to define the fitness of nanorobots. Subsequently, the MCDP can be seen as a procedure to search for the minimum values of the objective function:
 \begin{equation}
 \begin{split}
& \min \limits_{\vec{x}\in \mathcal{P}} \ o \left(\vec{x};A\right)
 \end{split}
 \end{equation}
In the MCDP, an optimization or niching method aims to locate all possible $\vec{x}^\ast\in\mathcal{P}$ (i.e., identifying all the tumors), which result in the smallest possible objective values:
\begin{equation}
\begin{split}
& o \left(\vec{x}^\ast;A\right)\leq o\left(\vec{x};A\right),\forall{\vec{x}\in{\mathcal{P}}}.
\end{split}
\end{equation}
The mapped $\emph{o}$ values in the immediate vicinity of an $\vec{x}^\ast$ should be all equal or higher than $o\left(\vec{x}^\ast; A\right)$. The corresponding solution set is denoted as $\mathcal{S}$.\par
More specifically, $\emph{N}$ nanorobotic agents
 $\emph{A}_1$,~$\emph{A}_2$,~$\cdots$,~$\emph{A}_N$
 are injected into the capillary around the high-risk tissue as shown in Fig. 1. The agents are employed to search for the solution set $\mathcal{S}$, with their initial locations $\vec{x}_1(t_0)$, $\vec{x}_2(t_0)$, $\cdots$,  $\vec{x}_N(t_0)$ at the initial time $t_0$. The initial speed of $\emph{A}_n$ $(n=1,2,...,N)$ is $v_n(\vec{x}_n(t_0))$, which is assumed to be equal to the blood flow velocity at $\vec{x}_n(t_0)$, as blood has a low Reynolds coefficient and the inertia force of agents can be ignored compared to the viscosity force \cite{sefidgar2015numerical}. The location of agent $\emph{A}_n$ can be updated according to Eq. (4) as the capillary network of the healthy tissue is grid-like\cite{chen2017touchable}:

 \begin{equation}
\begin{split}
\vec{x}_n(t_{k+1})&=\vec{x}_n(t_k)+d_n(t_k,t_{k+1})\vec{u}_{\angle{\phi_F(\vec{x}_n(t_k))+\phi_P(\vec{x}_n(t_k))}}\\
&=\vec{x}_n(t_k)+\sum_{l=1}^{L_n(t_k,t_{k+1})}v_n(\vec{x}_n(t_k))\delta t_{n,l} (t_k,t_{k+1})\\
   & \quad \times \vec{u}_{\angle{\phi_{n,l}(t_k,t_{k+1})}}.
\end{split}
\end{equation}

\begin{figure} [!htp]
\begin{center}
\epsfig{file=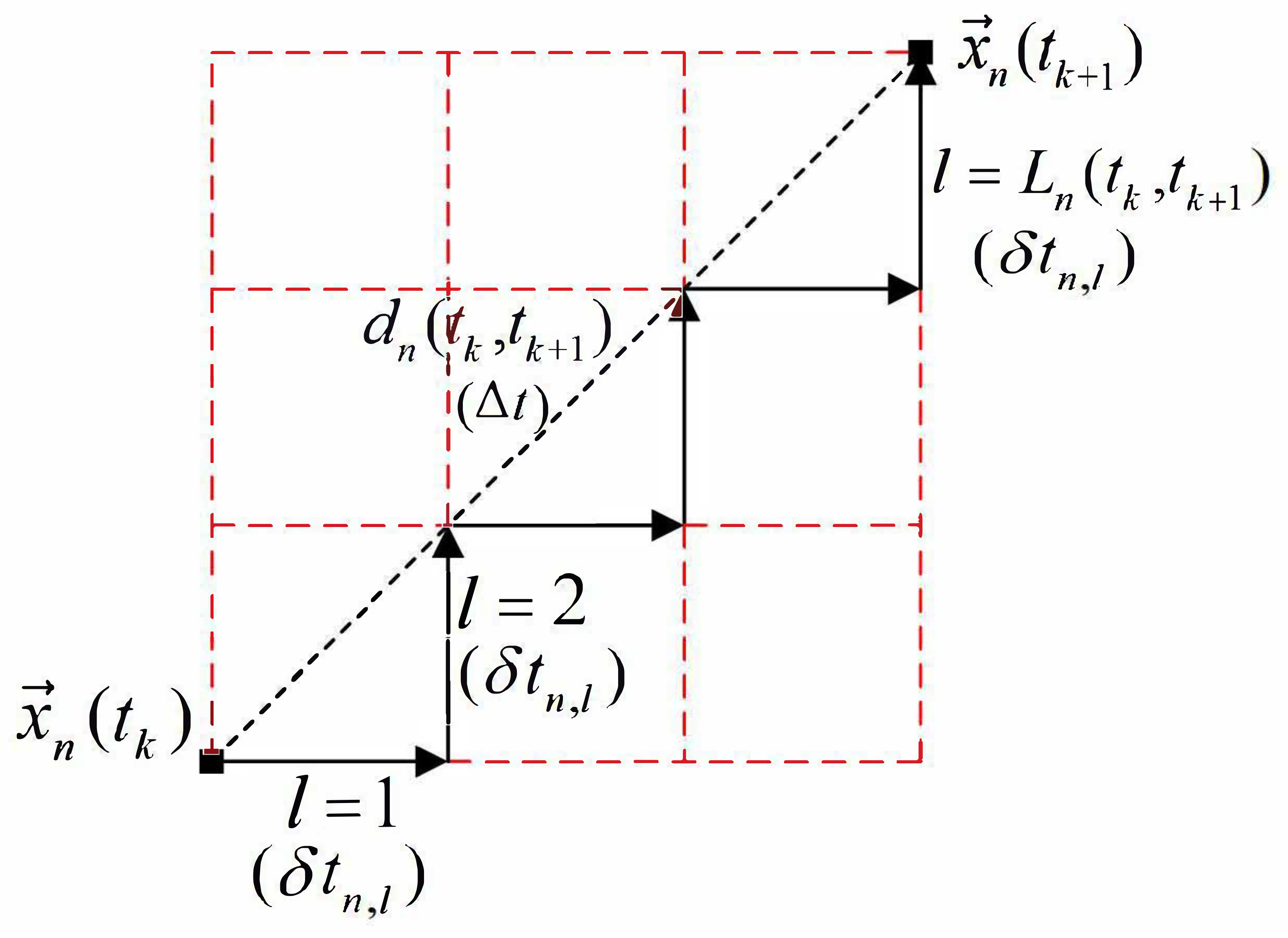,width=0.8\linewidth}
\end{center}
    \caption{Update of the location of agent $\emph{A}_n$ in one iteration}
    \label{fig:Fig.2}
\end{figure}
As illustrated in Fig. 2, when $\emph{A}_n$ does not meet the target, it will move continuously from $\vec{x}_n(t_k)$ at time $t_k$ to $\vec{x}_n(t_{k+1})$ at time $t_{k+1}$ in one iteration. The path of the nanorobot is composed of some linear segments. The term $d_n(t_k,t_{k+1})$ in (4) is the displacement length of $A_n$ from $\vec{x}_n(t_k)$ to $\vec{x}_n(t_{k+1})$, $L_n(t_k,t_{k+1})$ is the number of piecewise-linear sections along the trajectory of $\emph{A}_n$ from $\vec{x}_n(t_k)$ to $\vec{x}_n(t_{k+1})$ in the search space; the traveling time of the $l \rm^{th}$ linear section is $\delta t_{n,l} (t_k,t_{k+1})$ that satisfies $\sum_{l=1}^{L_{n}(t_k,t_{k+1})}\delta t_{n,l}(t_k,t_{k+1})=t_{k+1}-t_{k}=\triangle t$, where $\triangle t$ is the iteration period  of the external steering field; $\vec{u}_{\angle \phi}$ is a unit vector with angle $\phi$. The objective functions obtained by all agents are computed according to (1). When $\emph{A}_n$ meets the solution set $\mathcal{S}$ (i.e., a tumor is detected by a nanorobot), the stopping criteria are met. Otherwise, $\emph{A}_n$ will continue to move according to (4). A new agent will be randomly deployed within the injected area to ensure a constant number of agents in the medium, if an old agent  fully degenerates\cite{martel2009mri}.
\subsection{Landscape of Parameter Space}
The blood flow rate in a vascular network is proportional to the pressure difference between the arterial and venous sides and inversely proportional to the viscous and geometric resistances. Microvascular pressures in the arterial side are nearly equal in tumor and nontumorous vessels. Pressures in venular vessels are significantly lower in a tumor than those in a nontumorous tissue \cite{jain1988determinants}. The presence of tumors can thus affect the distribution of blood flow velocity in the capillaries of the high-risk tissue. The blood flow velocity was shown to decrease significantly when blood enters the tumors (dropping from $183.4\pm 35.0 \rm \mu m/s$ to $114.1\pm 26.1 \rm \mu m/s$) and then soon return to the prior level upon flowing out of the tumors \cite{wang2009blood}. Furthermore, the blood flow near malignant tumors has been studied using radioactive  $\rm ^{90}Y$ microspheres while treating patients. The qualitative investigation demonstrates a normal tissue flow/tumor flow ratio in the range of 3/1 to 30/1. In addition, the peripheral tumor areas show a higher flow rate than more central portions \cite{ellegala2003imaging,vaupel1989blood}.\par
Based on the aforementioned phenomena, we consider a representative artificial landscape to evaluate the performance of MCDP, which synthesizes the scenario of the blood flow variation around the tumor foci and can be regarded as the biological cost function to be optimized. In other words, the landscape shown in Fig. 3 can be used to represent $f(\vec{x})=o(\vec{x};\emph{A})(\rm mm/s)$, where $\vec{x}=(x,y)$ is in $\rm mm$ and the parameter range is $0\leq x,y \leq 10 \rm mm$.\par
As shown in Fig. 3, the landscape with three different humps represents the situation that there are three tumor foci. The blood flow velocity  decreases with the distance from the centers of the three tumors, (5.1, 7.3), (4.4, 5.9) and (5, 2.8) respectively. The objective function is expressed as follows\cite{engelbrecht2006fundamentals}.
\begin{equation}
\begin{split}
&f\left(\vec{x}\right)=\left \{
\begin{array}{ll}
 0, ~~~~~~~~~~~~~\hbox{$\sqrt{\left(x-5.1\right)^2+\left(y-7.3\right)^2}\leq 0.1\sqrt{3}$}\\
 0, ~~~~~~~~~~~~~\hbox{$\sqrt{\left(x-4.4\right)^2+\left(y-5.9\right)^2}\leq 0.1\sqrt{4}$}\\
 0, ~~~~~~~~~~~~~~~\hbox{$\sqrt{\left(x-5\right)^2+\left(y-2.8\right)^2}\leq 0.1\sqrt{5}$}\\
-\left|f_{1}\left(\vec{x}\right)+f_{2}\left(\vec{x}\right)+f_{3}\left(\vec{x}\right)\right|+0.18,~~~~\hbox{Otherwise}
\end{array} \right.
\end{split}
\end{equation}
where,
\begin{equation}
\begin{split}
f_{1}\left(\vec{x}\right)=\left(6-x\right)^2 \rm {exp}\left[-\left(\emph{x}-5\right)^2-\left(\emph{y}-6\right)^2\right]
\end{split}
\end{equation}
\begin{equation}
\begin{split}
f_{2}\left(\vec{x}\right)&=-0.1\left[x\slash 5-\left(x-5\right)^3-\left(y-5\right)^5-1\right]^2 \\
       & \quad \times \rm exp\left[-\left(\emph{x}-5\right)^2-\left(\emph{y}-5\right)^2\right]
\end{split}
\end{equation}
\begin{equation}
\begin{split}
f_{3}\left(\vec{x}\right)=0.5 \rm exp\left[-\left(\emph{x}-4\right)^2-\left(\emph{y}-5\right)^2\right]
\end{split}
\end{equation}

\begin{figure} [!htp]
\centering

\epsfig{file=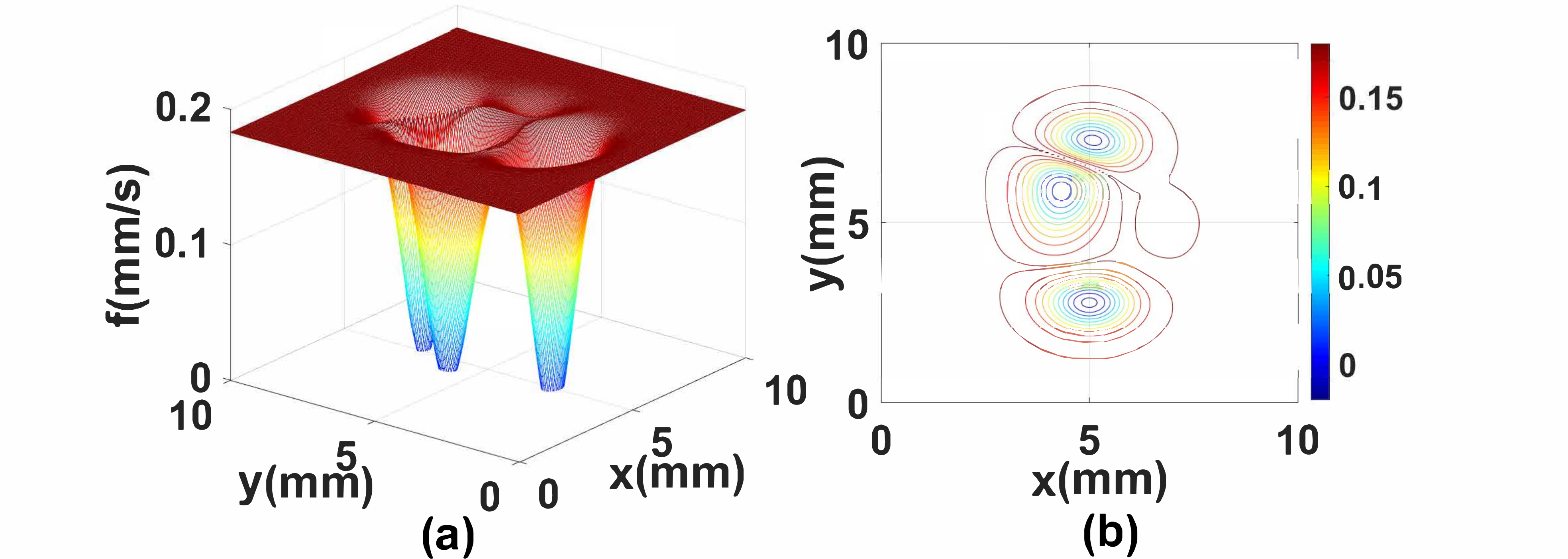,width=1.1\linewidth}

    \caption{ (a) Three-dimensional plot of the artificial landscape, and (b) blood velocity contour of the parameter space.}
    \label{fig:Fig.3}

\end{figure}

\subsection{Capillary Network Model of Tumors}

In tumor vasculature, the large surface areas of capillaries make them the ideal site to provide nutrients and oxygen for the growing of tumors \cite{rieger2015integrative}. For tumors, angiogenesis appears to be a critical determinant of growth, invasion, and metastatic potential \cite{zetter1998angiogenesis, weidner1991tumor, folkman1971tumor}. The angiogenic activity is located within a few hundred micrometers from the tumor rim. Fueling further growth, the resulting neovasculature is progressively coopted together with the original blood vessels by the expanding tumor mass while also pushing the neovascularization zone further into normal tissue \cite{welter2013interstitial}. The hastily formed new tumor microcirculation is always different from the host vasculature due to the influence of tumor angiogenesis factors and reduction of the available oxygen and nutrients \cite{jain1988determinants, vaupel1987blood, anderson1998continuous, song1984effect}. Hence tumors are known to contain many tortuous vessels, shunts, vascular loops, widely variable intervascular distances, and large avascular areas, while normal capillaries are almost uniformly distributed to ensure adequate oxygen and nutrients transportation through the tissue \cite{gazit1995scale}. The microvascular density is greater in the periphery of the highly vascularized tumor than that of the surrounding normal tissue \cite{lee2006flow}. These characteristics of tumor vasculature are well synthesized by the invasion percolation method, which is a classical statistical growth process governed by local substrate properties \cite{baish1996role}.\par
The simulation of vascular growth begins with a square lattice of discrete points that represents potential paths of vascular growth. Invasion percolation is implemented by first assigning uniformly distributed random values or strengths to each point on the underlying lattice. Starting at an arbitrary point the network occupies the lattice point adjacent to the current network that has the lowest strength. Growth is iterated until the desired lattice occupancy is reached. Blood vessels are assumed to connect all adjacent occupied lattice points. Blood is supplied at the starting point and withdrawn from the point nearest to the opposite corner. The network is then pruned to retain only those parts of the network with nonzero flow, leaving what is known as the ``backbone'' of the percolation cluster. Fig. 4 depicts the invasion percolation result after  260 growth steps, which also shows the backbone of the percolation clusters after pruning the vessels with zero blood flow.\par 

\begin{figure} [!htp]
\centering
\epsfig{file=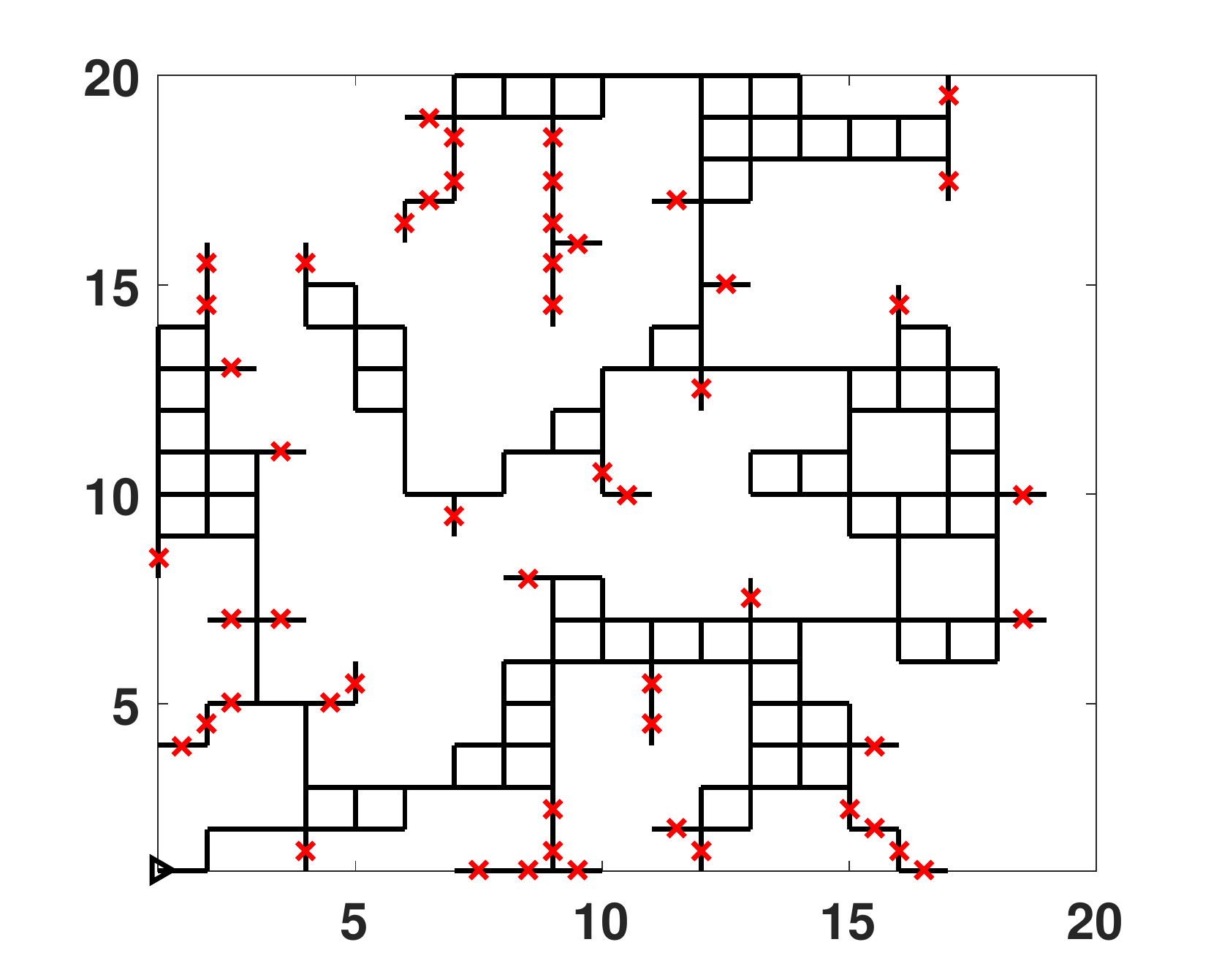,width=0.7\linewidth}

    \caption{Computer-generated percolation network after 260 growth steps of occupancy after pruning of vessels with zero blood flow.}
    \label{fig:Fig.4}
\end{figure}

It is a common strategy to restrict the vessels to run only parallel to the coordinate axes in modeling microvasculature \cite{anderson1998continuous,lee2006flow,jackson2011modeling,baish2000fractals}. Subsequently, it is appropriate to model the tumor capillary network using the latticed model with a fixed microvascular density for normal tissues and a greater microvascular density for the periphery of tumors. In this way, we simulate the tumor-induced angiogenesis for the blood flow velocity profile in (5) as shown in Fig. 5. The global capillary network is a square with an area of $10\rm mm\times 10mm$. There are three different tumor foci T1, T2, and T3, which are represented by three different square areas. The center coordinates of the tumors are (5.1, 7.3), (4.4, 5.9), (5, 2.8) and the side lengthes are 0.8mm, 1mm, 1.2mm, respectively. The centers of the tumor foci where the velocity of nanorobots becomes zero are three circular areas with a radius of $100 \rm \mu m$. We define the intercapillary distances to be $100 \rm \mu m$ for the normal capillary network and $50\rm \mu m$ for capillaries in the periphery of the tumors. The periphery we define here corresponds to the blood velocity contour with the value of $60 \rm \mu m/s$. The level of occupancy on the lattice is chosen to be 60\% before pruning \cite{baish1996role}. For each vascular element in Fig. 5, the flow rate is assumed to follow the Poiseuille's law. A number of capillary elements come together at each node, which satisfies mass conservation and therefore the sum of all flows at each node is zero. The general direction of blood flow is from the bottom left coordinate (0, 0) to the top right coordinate (10, 10), which is $45^{\circ}$ to the $x$ and $y$ axes, respectively as shown in Fig. 5. \par

\begin{figure} [!htp]
\centering
\epsfig{file=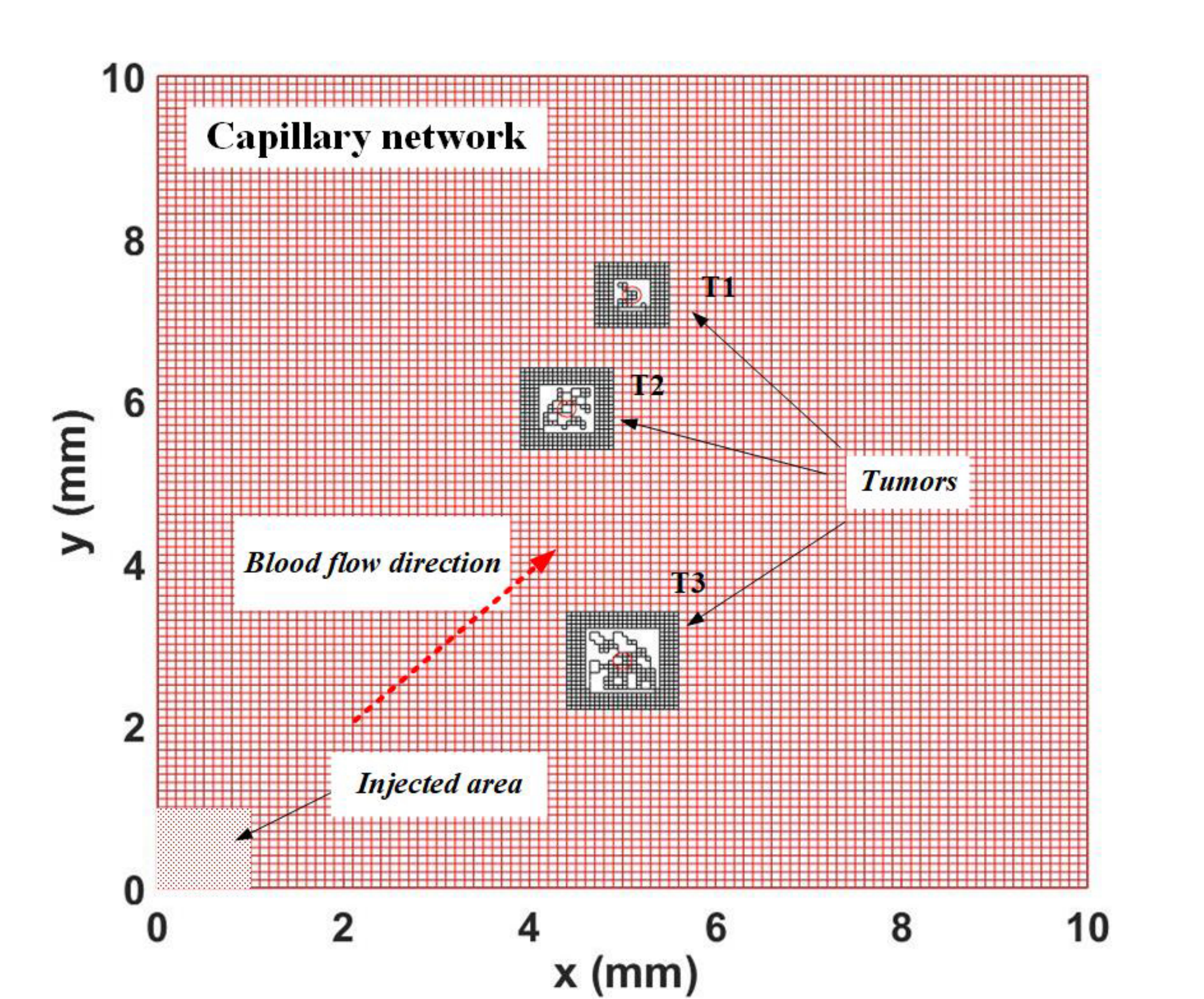,width=1.05\linewidth}
    \caption{Simulated vascular network for the blood flow velocity profile in (5). The vasculature is created by regular space-filling arrays with different microvascular densities representing normal and peritumoral tissues and invasion-percolation-based structure representing tumor tissues with a reduced microvascular density. The tumor foci centers are denoted by the red circles.}
    \label{fig:Fig.5}
\end{figure}

\section{NGA-inspired MCDP}
The GA firstly proposed and analyzed by J. H. Holland in 1975 is commonly used to solve optimization problems\cite{de2006evolutionary,mitchell1998introduction, holland1992adaptation}. As an effective method for solving complex problems, once the genetic representation and the fitness function are defined, a GA proceeds to initialize a population of solutions and then to improve it through repetitive application of the mutation, crossover, inversion and selection operators \cite{bies2006genetic}.\par
However, traditional GAs will push an artificial population towards convergence though natural evolutionary processes maintain a variety of species with each occupying a separate ecological niche. In nature, a niche is viewed as an organism's task in the environment, and a species is a collection of organisms with similar features. The subdivision of environment on the basis of an organism's role helps stable subpopulations to form around different niches in the environment \cite{deb1989investigation}. Niching technique has recently been introduced into GAs to make it possible to find more than one local optimum of a multimodal landscape. Niching technology can make individuals evolve in a special environment by adjusting the fitness of individuals and the replacement strategy to generate the new generation. This method can maintain diversity of evolution population and acquire multiple solutions at the same time. Three general niching techniques: pre-selection, crowding, and fitness sharing are popular. The NGA is widely used in the multimodal and non-monotonic function optimization. In NGA, a niche is commonly referred to as the location of each optimum in the search space, and the fitness represents the resource of that niche. The organisms in a niche can be defined as similar individuals in terms of similarity metrics. The pseudo code of NGA is given in Algorithm 1.\\
\begin{algorithm}
 \caption{NGA}
  \begin{algorithmic}[1]
   \State Choose an initial population  $\emph{A}_1$,~$\emph{A}_2$,~$\cdots$,~$\emph{A}_N$
   with their initial locations $\vec{x}_1^0$, $\vec{x}_2^0$, $\cdots$, $\vec{x}_N^0$
   \State Evaluate fitness of agents $\emph{A}_1$,~$\emph{A}_2$,~$\cdots$,~$\emph{A}_N$.
   \For {$k$ less than the initialized iteration number $\emph{K}$}
    \State Use normal genetic operators to get new positions $\vec{x}_1^{(k+1)}$, $\vec{x}_2^{(k+1)}$, $\cdots$, $\vec{x}_N^{(k+1)}$ of new agents
    \State Replace the old agents (positions $\vec{x}_1^k$, $\vec{x}_2^k$, $\cdots$, $\vec{x}_N^k$) with new agents (positions $\vec{x}_1^{(k+1)}$, $\vec{x}_2^{(k+1)}$, $\cdots$, $\vec{x}_N^{(k+1)}$)
    \State Use niching technology in agents:
    \While {$Distance(\vec{x}_i^{(k+1)}, \vec{x}_j^{(k+1)})\leq \emph{L}$}
    \State Penalize the worse agent
    \EndWhile
    \State Evaluate fitness of agents (the initial population of next generation) at positions
    $\vec{x}_1^{(k+1)}$, $\vec{x}_2^{(k+1)}$, $\cdots$, $\vec{x}_N^{(k+1)}$
    \EndFor
  \end{algorithmic}
\end{algorithm} \\

As for the selection operator, we use the method of roulette wheel selection, which is the first selection method developed by Holland [44]. The probability of selection for each individual $i$, $\emph{P}_i$, is defined by:
 \begin{equation}
\begin{split}
\emph{P}_i=\frac{F_i}{\sum_{j=1}^{N}F_j}
\end{split}
\end{equation}
where $F_i$ is equal to the fitness of individual $i$. As for the crossover operator, we use traditional two-parent crossover operator and characterize the result of this process using real-value coding. The $N$ individuals are paired at random, yielding $N/2$ couples. For each couple, crossover may or may not occur. Crossover does not occur with the probability of $1-p_c$. Subsequently, both individuals proceed to the mutation stage. Otherwise, the couple generates two children \emph{via} crossover, and only the children continue to the mutation stage.  The crossover and mutation processes are as follows:
\begin{equation}
\begin{split}
\left\{
\begin{array}{c}
  \vec{x}_n^{(k+1)}=\alpha\vec{x}_n^k+(1-\alpha)\vec{x}_{n+1}^k \\
  \vec{x}_{n+1}^{(k+1)}=\alpha\vec{x}_{n+1}^k+(1-\alpha)\vec{x}_n^k
\end{array}\right.
\end{split}
\end{equation}

\begin{equation}
\begin{split}
\vec{x}_n^{(k+1)}=\vec{x}_n^k+\sigma
\end{split}
\end{equation}
where $\alpha$ follows a uniform distribution between 0 and 1 (i.e. $\alpha\sim U(0,1)$), and $\sigma$ follows a Gaussian distribution with mean 0 and variance 1 (i.e. $\sigma\sim N(0,1)$). Then, we use a niching method with modified crowding scheme to preserve population diversity. In crowding, each individual is compared to its neighbor within a constant distance, and the worse one is punished with a less opportunity to produce an offspring. Inspired by the physiological viewpoint that offspring resulting from inbreeding tends to be unhealthy, we further propose a modified NGA. In the algorithm, the initial population is separated into two groups with the same number of individuals randomly; each individual in one group can only perform crossover with an arbitrary individual in another group; the rest of the processes are similar with the original NGA. The modified NGA is represented in the pseudo code of Algorithm 2.
\begin{algorithm}
 \caption{Modified NGA}
  \begin{algorithmic}[1]
   \State Choose two initial populations  $\emph{A}_1$,~$\emph{A}_2$,~$\cdots$,~$\emph{A}_M$ and $\emph{A}_{M+1}$,~$\emph{A}_{M+2}$,~$\cdots$,~$\emph{A}_N$
   with their initial locations $\vec{x}_1^0$, $\vec{x}_2^0$, $\cdots$, $\vec{x}_M^0$ and $\vec{x}_{M+1}^0$, $\vec{x}_{M+2}^0$, $\cdots$, $\vec{x}_N^0$
   \State Evaluate fitness of agents $\emph{A}_1$,~$\emph{A}_2$,~$\cdots$,~$\emph{A}_N$.
   \For {$k$ less than the initialized iteration number $\emph{K}$}
    \State Use normal genetic operators with crossover between two groups to get new positions $\vec{x}_1^{(k+1)}$, $\vec{x}_2^{(k+1)}$, $\cdots$, $\vec{x}_M^{(k+1)}$ and $\vec{x}_{M+1}^{(k+1)}$, $\vec{x}_{M+2}^{(k+1)}$, $\cdots$, $\vec{x}_N^{(k+1)}$ of new agents.
   \State Replace the old agents (positions $\vec{x}_1^k$, $\vec{x}_2^k$, $\cdots$, $\vec{x}_M^k$ and $\vec{x}_{M+1}^k$, $\vec{x}_{M+2}^k$, $\cdots$, $\vec{x}_N^k$) with new agents (positions $\vec{x}_1^{(k+1)}$, $\vec{x}_2^{(k+1)}$, $\cdots$, $\vec{x}_M^{(k+1)}$ and $\vec{x}_{M+1}^{(k+1)}$, $\vec{x}_{M+2}^{(k+1)}$, $\cdots$, $\vec{x}_N^{(k+1)}$).
   \State Use niching technology in agents:
   \While {$Distance(\vec{x}_i^{(k+1)}, \vec{x}_j^{(k+1)})\leq \emph{L}$}
   \State Penalize the worse agent
   \EndWhile 
    \State Evaluate fitness of agents (the initial population of next generation) at positions
    $\vec{x}_1^{(k+1)}$, $\vec{x}_2^{(k+1)}$, $\cdots$, $\vec{x}_N^{(k+1)}$
    \EndFor
  \end{algorithmic}
\end{algorithm}

In the following, we will formulate the MCDP based on the similar evolutionary mechanism with the aforementioned NGAs but taking into account realistic \emph{in vivo} propagation, controlling, and tracking scenarios.

Nanorobots loaded with contrast medium molecules are injected into the tissue region. The injection area is a $1 \rm mm\times1mm$ square as shown in Fig. 5. Each nanorobot adjusts its own location based on the comparison with its offspring location under the influence of the blood flow and the external magnetic field. Fig. 6 gives the movement direction of agent $A_n$ in MCDP. The location of $A_n$ at step $k$ is denoted by the vector $\vec{x}_n^{k}$; the ideal child location of $A_n$ following the NGA is denoted by the vector $\vec{x}_n^{(k+1)}$ and the actual location of $A_n$ following the MCDP is denoted by the vector $\vec{x}_n^{k+1}$ ; The blood velocity, the nanorobot's velocity subject to the magnetic field without the influence of blood flow, and their resultant velocity are denoted by $\vec{v}_b$, $\vec{v}_m$, and $\vec{v}_c$, respectively. Subsequently, agent $A_n$ will move by one step in the direction of $\vec{v}_c$ and the new position of $A_n$ denoted by $\vec{x}_n^{k+1}$ will be obtained. $\vec{x}_n^{k+1}$ then plays the role of parent in the next generation. Following this process, each nanorobot updates its location step by step.  Note that the  capillary network of healthy tissue and the tumor capillary network used in the simulation procedure are discontinuous two-dimensional grids; therefore, all agents' locations are mapped to the locations associated with the nearest blood vessel.

Subsequently, agent $A_n$ evolves according to the following equations:
\begin{equation}
\begin{split}
\vec{u}_{\angle \phi (\vec{x}_n^k)}=\frac{\vec{x}_n^{(k+1)}-\vec{x}_n^k}{|\vec{x}_n^{(k+1)}-\vec{x}_n^k|}
\end{split}
\end{equation}

\begin{equation}
\begin{split}
\vec{x}_n^{k+1}=\vec{x}_n^k+\sum_{l=1}^{L_n}v_n\delta t_{n,l} \vec{u}_{\angle \phi_{n,l}}
\end{split}
\end{equation}

\begin{figure} [!htp]
\begin{center}

\epsfig{file=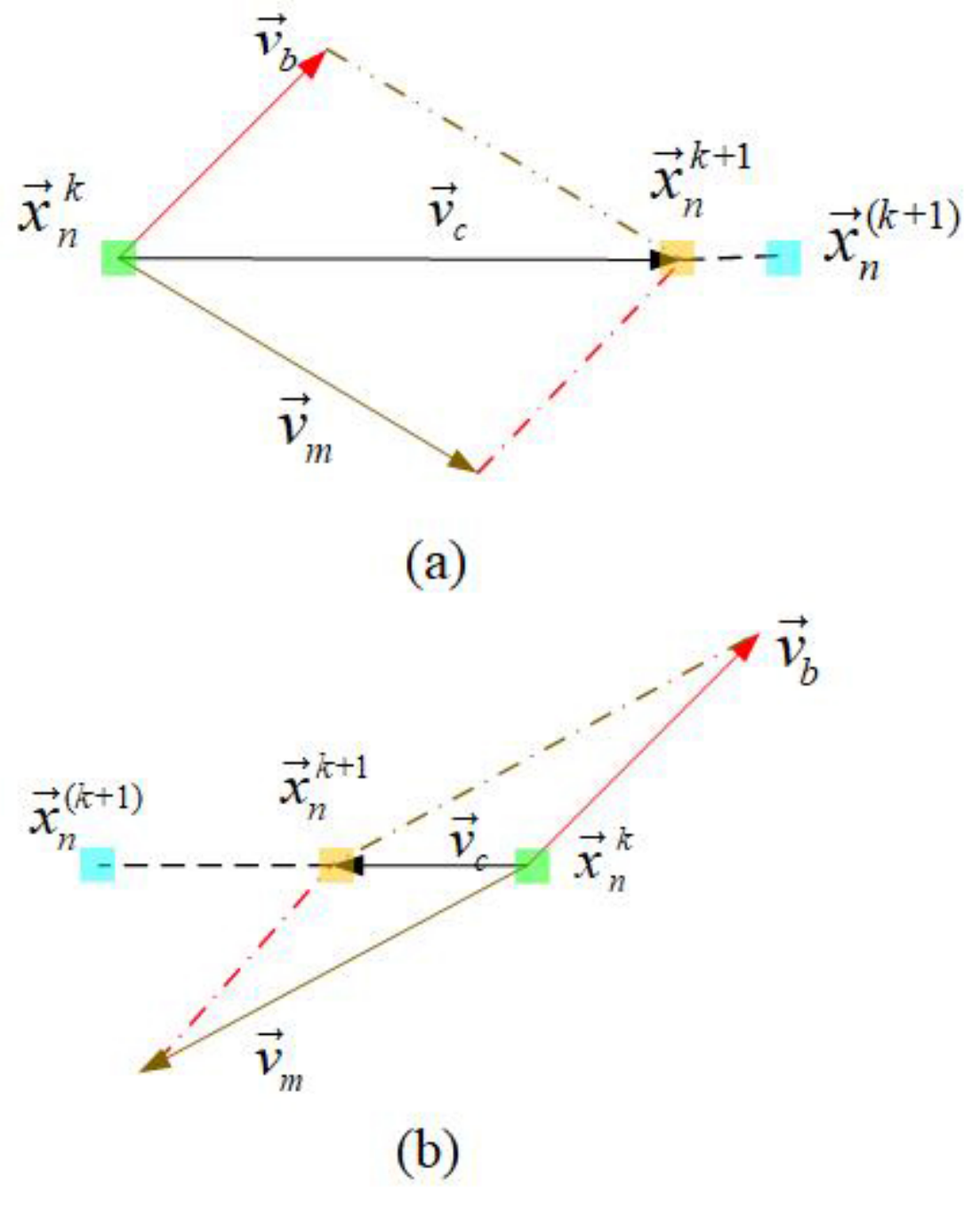,width=0.6\linewidth}

    \caption{Update of the location of a nanorobot: (a) the angle between the directions of the blood flow and the magnetic field is acute; and (b) the angle between the directions of the blood flow and the magnetic field is obtuse.}
    \label{fig:Fig.6}

\end{center}
\end{figure}

The index $n=1, 2, \cdots, N$; $\phi(\vec{x}_n)$ denotes the angle generated by the co-force of the external magnetic field and the blood flow on agent $A_n$ at $\vec{x}_n^k$, $\phi_{n,l}$  is the direction of the $l^{ \rm th}$ linear section, $\vec{u}_{\angle \phi}$ is a unit vector with angle $\phi$, $L_n$ is the number of piecewise-linear sections along the trajectory of the $n^{\rm th}$ agent in one iteration and $\delta t_{n,l}$ is the traveling time of the $l^{\rm th}$ linear section satisfying $\sum_{l=1}^{L_n}\delta t_{n,l}=\Delta t$ with $\Delta t$ being the constant interval of the controlling cycle. Eq. (12) is used to calculate the new movement direction of the agent $A_n$ by taking into the agent's current position and its offspring position obtained by the NGA operators. Eq. (13) is used to evaluate the updated position of the agent $A_n$ given the steering field and the underlying vascular network.

From the description above, we can see that the ideal location of each new agent provides a direction for the movement of its parent nanorobot. Every parent nanorobot swims in the direction of its offspring agent under the steering of the external magnetic field and the blood flow field. The key difference between the NGA-inspired MCDP proposed here and the NGA presented previously is the offspring used. For NGA, the offsprings can be at any random locations in the continuous searching space, while the offsprings produced in the MCDP are all in the locations of blood vessels as the nanorobots can only swim in the capillary network. The NGA-inspired MCDP is described in Algorithm 3.
\begin{algorithm}
 \caption{MCDP Algorithm}
  \begin{algorithmic}[1]
   \State Choose an initial population  $\emph{A}_1$,~$\emph{A}_2$,~$\cdots$,~$\emph{A}_N$
   with their initial locations $\vec{x}_1^0$, $\vec{x}_2^0$, $\cdots$, $\vec{x}_N^0$
   \State Evaluate fitness of agents $\emph{A}_1$,~$\emph{A}_2$,~$\cdots$,~$\emph{A}_N$.
      \For {$k$ less than the initialized iteration number $\emph{K}$}
    \State Use normal genetic operators to get new positions $\vec{x}_1^{(k+1)}$, $\vec{x}_2^{(k+1)}$, $\cdots$, $\vec{x}_N^{(k+1)}$ of new agents
   \While {the agents stopping criterion not met}
   \State Move agents towards the offspring locations $\vec{x}_1^{(k+1)}$, $\vec{x}_2^{(k+1)}$, $\cdots$, $\vec{x}_N^{(k+1)}$ with the same steps respectively and get the new positions of agents $\vec{x}_1^{k+1}$, $\vec{x}_2^{k+1}$, $\cdots$, $\vec{x}_N^{k+1}$
   \EndWhile
   \State Use niche technology in agents:
   \While {$Distance(\vec{x}_i^{k+1}, \vec{x}_j^{k+1})\leq \emph{L}$}
    \State Penalize the worse agent
   \EndWhile
    \State Evaluate fitness of agents (the initial population of next generation) at positions $\vec{x}_1^{k+1}$, $\vec{x}_2^{k+1}$, $\cdots$, $\vec{x}_N^{k+1}$.
    \EndFor
  \end{algorithmic}
\end{algorithm}

The initial deployment region of all the agents (i.e., the injected area shown in Fig. 5) is confined within a small area due to the practical constraint of the NGA-inspired MCDP, which is different from the normal NGA where all the agents are distributed in the whole searching space in the initial stage. All the nanorobots swim in the high-risk tissue under the external magnetic field to detect tumor areas using the MCDP given above. In case any nanorobot travels outside the search space, a new nanorobot will be deployed in the initial area to keep the number of nanorobots unchanged. The escaped nanorobot will degrade in the human body without causing any harmful effect \cite{chen2017touchable}.

Inspired by the modified NGA, we introduce the new crossover scheme into MCDP. In this scenario, the nanorobots employed are divided into two groups randomly chosen in the initial phase with the other processes being similar to the aforementioned MCDP algorithm.

\section{Performance Analysis}
We use several numerical examples to elaborate on the NGA-inspired MCDP. The blood flow velocity profile presented in Section II-B and the vascular network model presented in Section II-C are applied to synthesize tumor angiogenesis. In the simulation, 12 nanorobots are employed to execute the MCDP. The small number of nanorobots is chosen to minimize any side effect resulting from an excessive amount of nanorobots on the human body. The initial injection area is $0\leq x, y \leq 1 \rm mm$ as shown in Fig. 5. The initial speed of nanorobots is assumed to be $210 \rm \mu m/s$, which can be realized according to \cite{martel2009mri}. The period of iteration is set to be $4 \rm s$ and the number of iterations is 60. The probabilities of crossover and mutation are 0.9 and 0.1, respectively. The crowding distance is set to be $0.15 \rm mm$.

Fig. 7 shows a typical curve of minimum outcomes obtained from the agents over multiple iterations for the blood velocity profile in (5) by using the NGA-inspired MCDP. It can be seen that an agent with the best performance can detect a tumor successfully at the $22^{\rm th}$ iteration. To compare with the cancer detection method relying on systemic circulation without applying any metaheuristics under the same physical scenario, we also present the result obtained from the random searching strategy, which means the nanorobots swim in the capillary network without any external influence except the blood flow. In this case, we can see that the best agent can detect a tumor after 41 iterations, which is slower compared to the NGA-inspired MCDP. Furthermore, for comparison from the algorithmic perspective, we show the result for the standard NGA without any physical constraints. In this case, a target can be detected after 4 iterations.
\begin{figure} [!htp]
\begin{center}

\epsfig{file=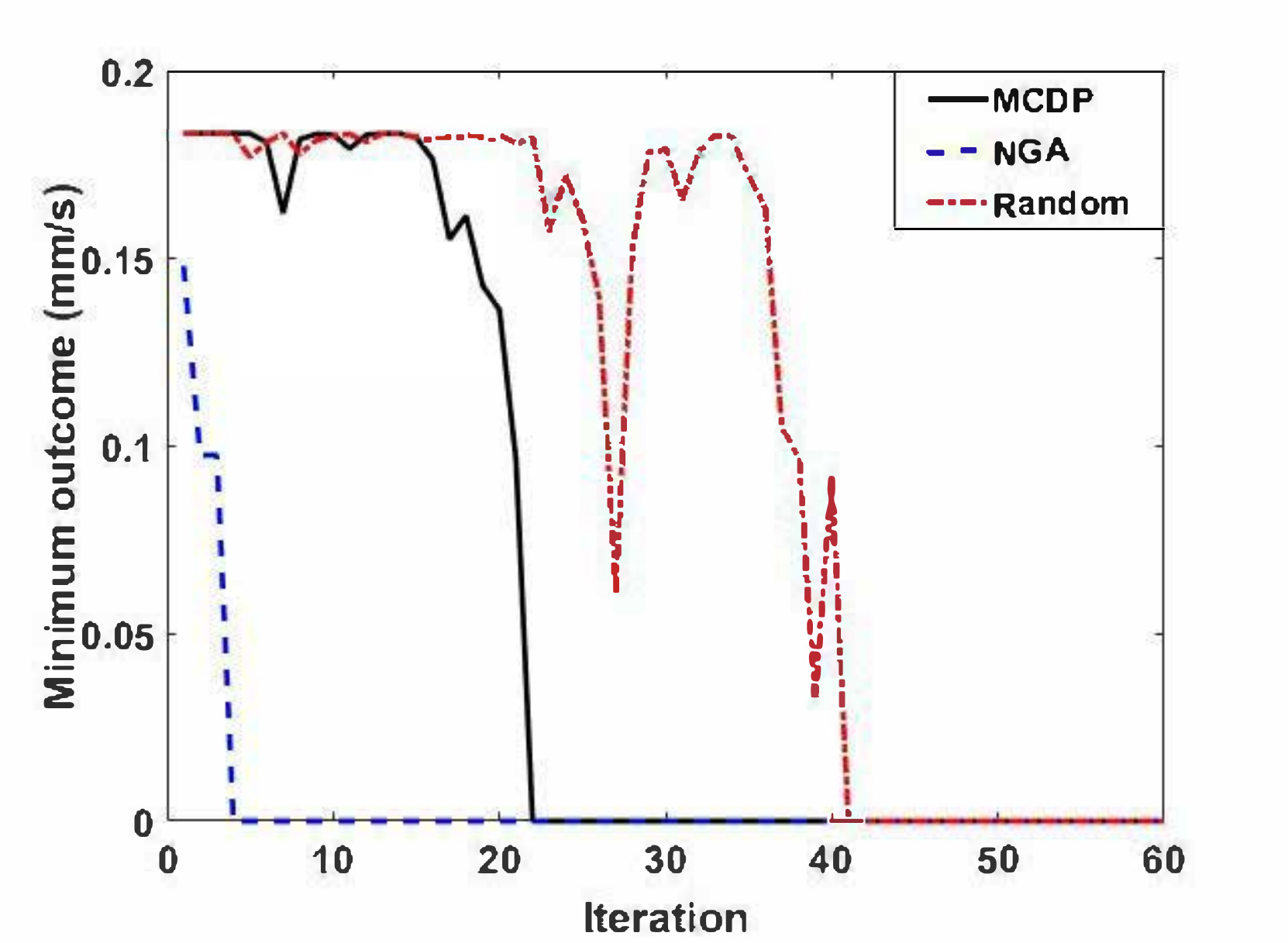,width=1.0\linewidth}

    \caption{Minimum outcomes obtained over iterations.}
    \label{fig:Fig.7}

\end{center}
\end{figure}

Fig. 8 shows the final locations of agents after 60 iterations by using the three different algorithms. We can see that the agents in the NGA and MCDP algorithms have a good performance of clustering with nearly all the agents swarming around the different target areas. On the contrary, the agents in the random searching strategy are scattered in the searching space without any swarming behavior.

\begin{figure} [!htp]
\begin{center}

\epsfig{file=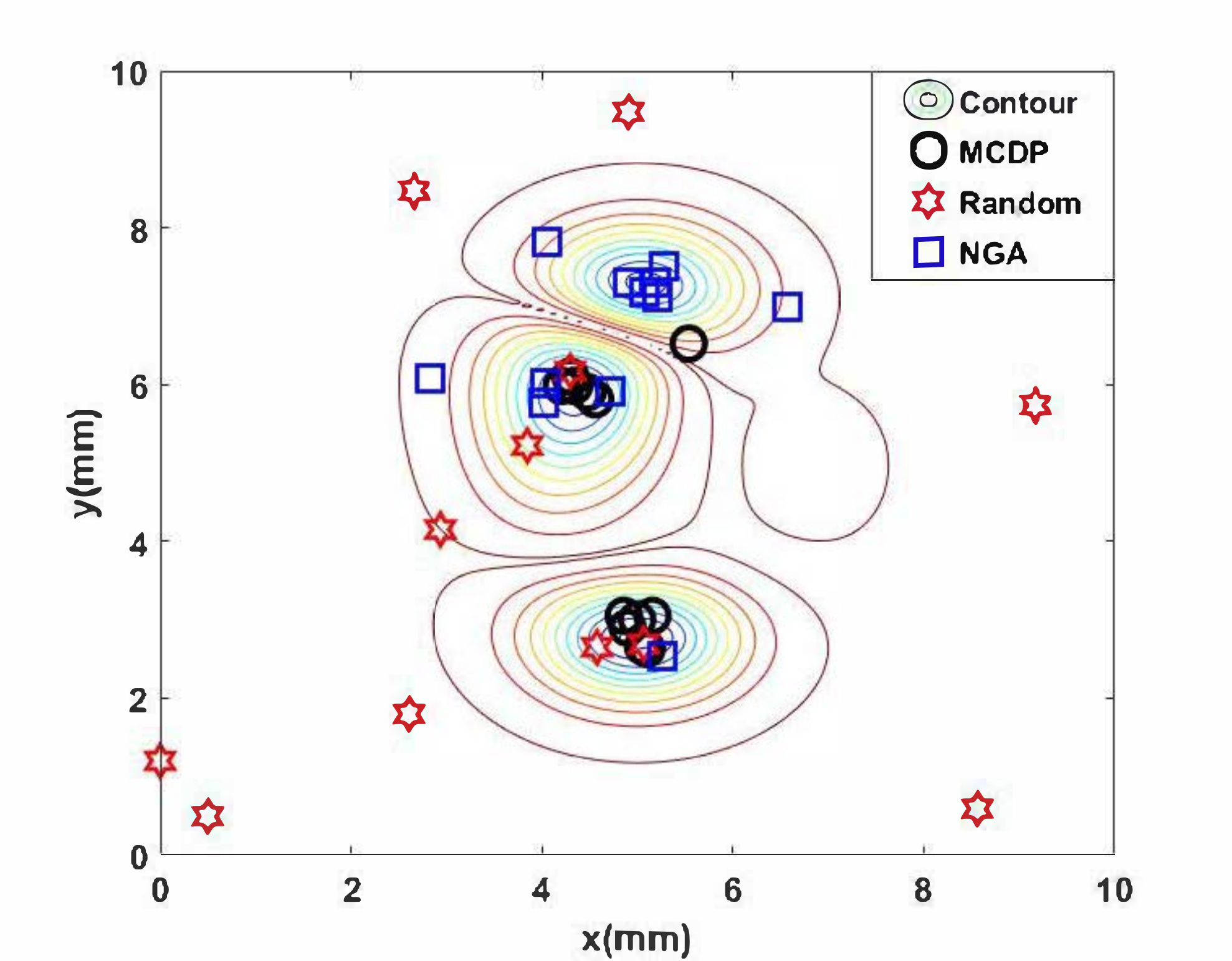,width=1.0\linewidth}

    \caption{Final positions of agents in the searching space.}
    \label{fig:Fig.8}

\end{center}
\end{figure}

To provide statistical analysis of robustness and precision of the NGA-inspired MCDP, we have carried out 500 independent simulation runs. Fig. 9(a), (b), and (c) show the histograms of the quantity of agents located at the targets using the MCDP, random searching and normal NGA strategies with the means of 1.54, 0.07 and 3.85, and the variances of 1.68, 0.08 and 1.79, respectively. It is evident that the proposed MCDP significantly outperforms random searching in term of the number of agents that can detect the cancer foci successfully. The performance of the MCDP strategy is not as good as the standard NGA, which is due to the fact that the MCDP is subject to various physical constraints.
\begin{figure} [!htp]
\begin{center}

\epsfig{file=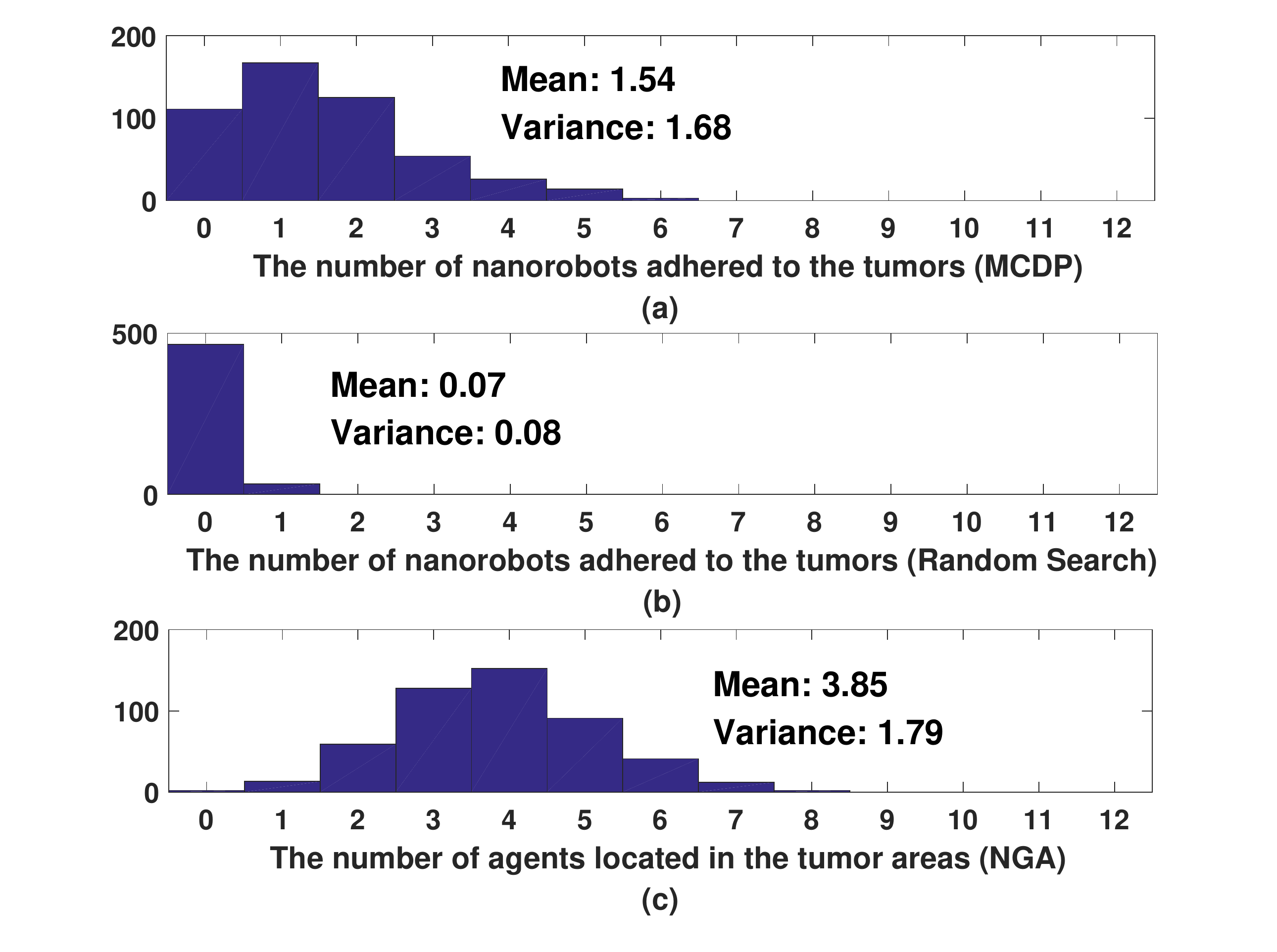,width=1.0\linewidth}

    \caption{ (a) Histogram of the quantity of agents located at the tumor areas using MCDP strategy. (b) Histogram of the quantity of agents located at the tumor areas using random searching strategy. (c) Histogram of the quantity of agents located at the target areas using normal NGA.}
    \label{fig:Fig. 9}

\end{center}
\end{figure}

Fig. 10 shows the curves of minimum outcomes obtained from the agents over multiple iterations by using the modified NGA, random searching and MCDP algorithms. Comparing Fig. 10 to Fig. 7, it can be seen that the MCDP inspired by the modified NGA performs worse than the one derived from the standard NGA in terms of the time spent in detecting a single tumor because of the decrease in exploitation for the modified NGA. On the other hand, the other comparison results between the three methods are similar with the earlier observations.

\begin{figure} [!htp]
\begin{center}

\epsfig{file=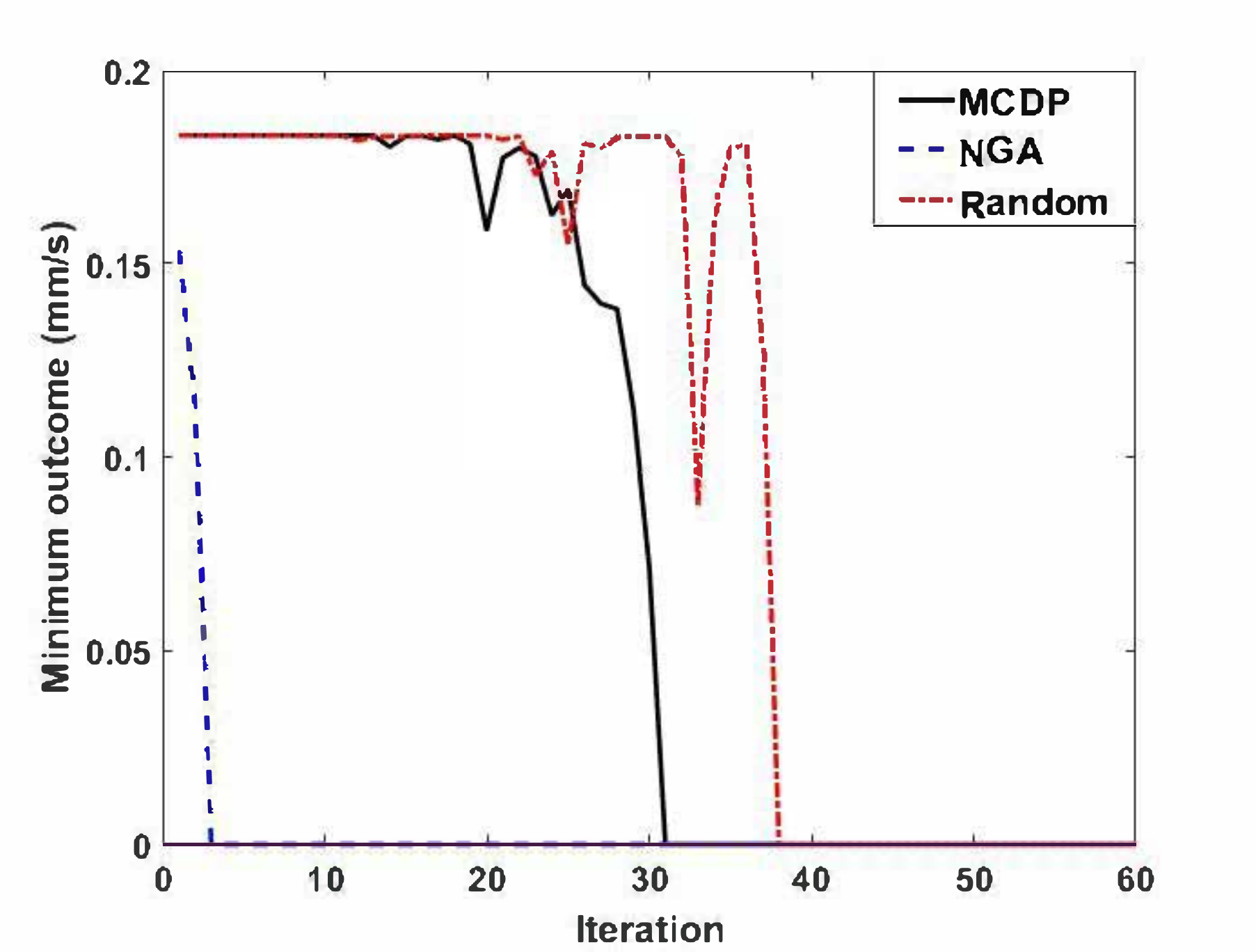,width=1.0\linewidth}

    \caption{Minimum outcomes obtained over iterations.}
    \label{fig:Fig. 10}

\end{center}
\end{figure}

Fig. 11 shows the final locations of agents after 60 iterations in the second case. It can be seen that the result in Fig. 11 is similar with that in Fig. 8.
\begin{figure} [!htp]
\begin{center}

\epsfig{file=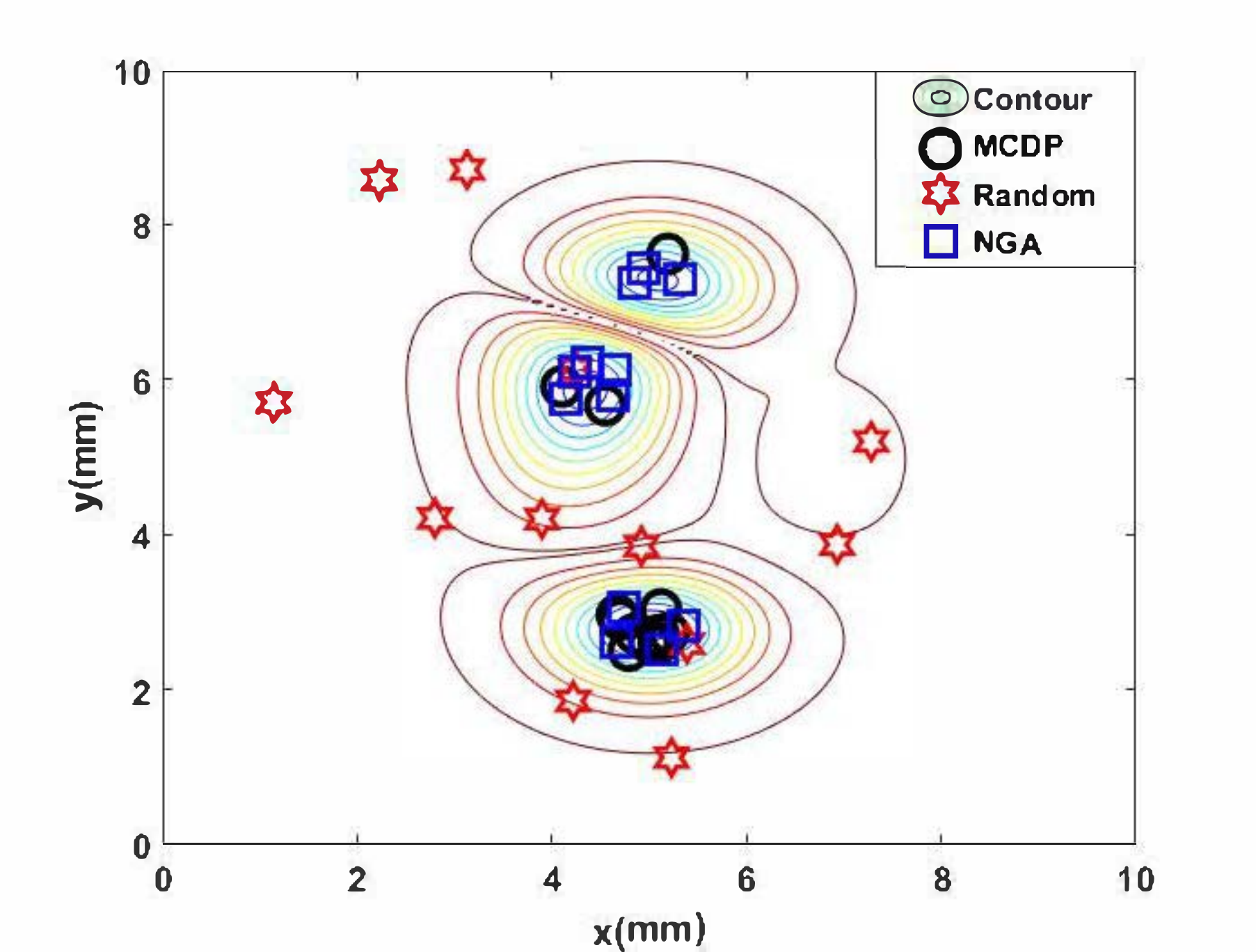,width=1.0\linewidth}

    \caption{Final positions of agents in the searching space.}
    \label{fig:Fig.11}

\end{center}
\end{figure}

Fig. 12 gives the histogram of the quantity of agents located at the tumor foci using the three algorithms in the second case. Though the observations made in Fig. 12 are somewhat similar with those from Fig. 9, it is interesting to note that the mean in the second case for MCDP is 2.40, which is greater than the value of 1.54 in Fig. 9(a) due to the increase in exploration for the modified NGA.
\begin{figure} [!htp]
\begin{center}
\epsfig{file=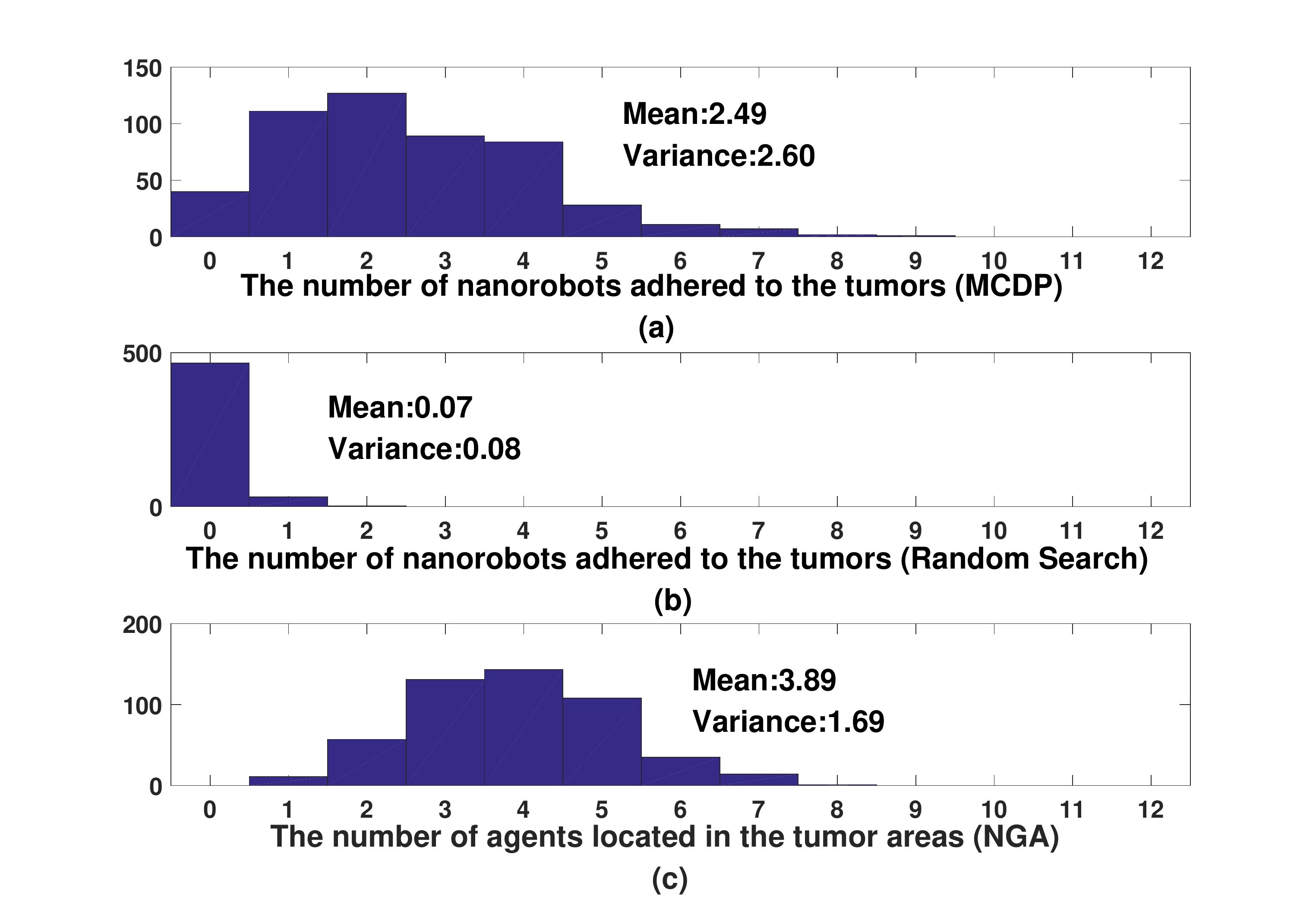,width=1.0\linewidth}
    \caption{ (a) Histogram of the quantity of agents located at the tumor areas using MCDP strategy. (b) Histogram of the quantity of agents located at the tumor areas using random searching strategy. (c) Histogram of the quantity of agents located at the target areas using modified NGA}
    \label{fig:Fig.12}
\end{center}
\end{figure}

\begin{figure} [!htp]
\begin{center}

\epsfig{file=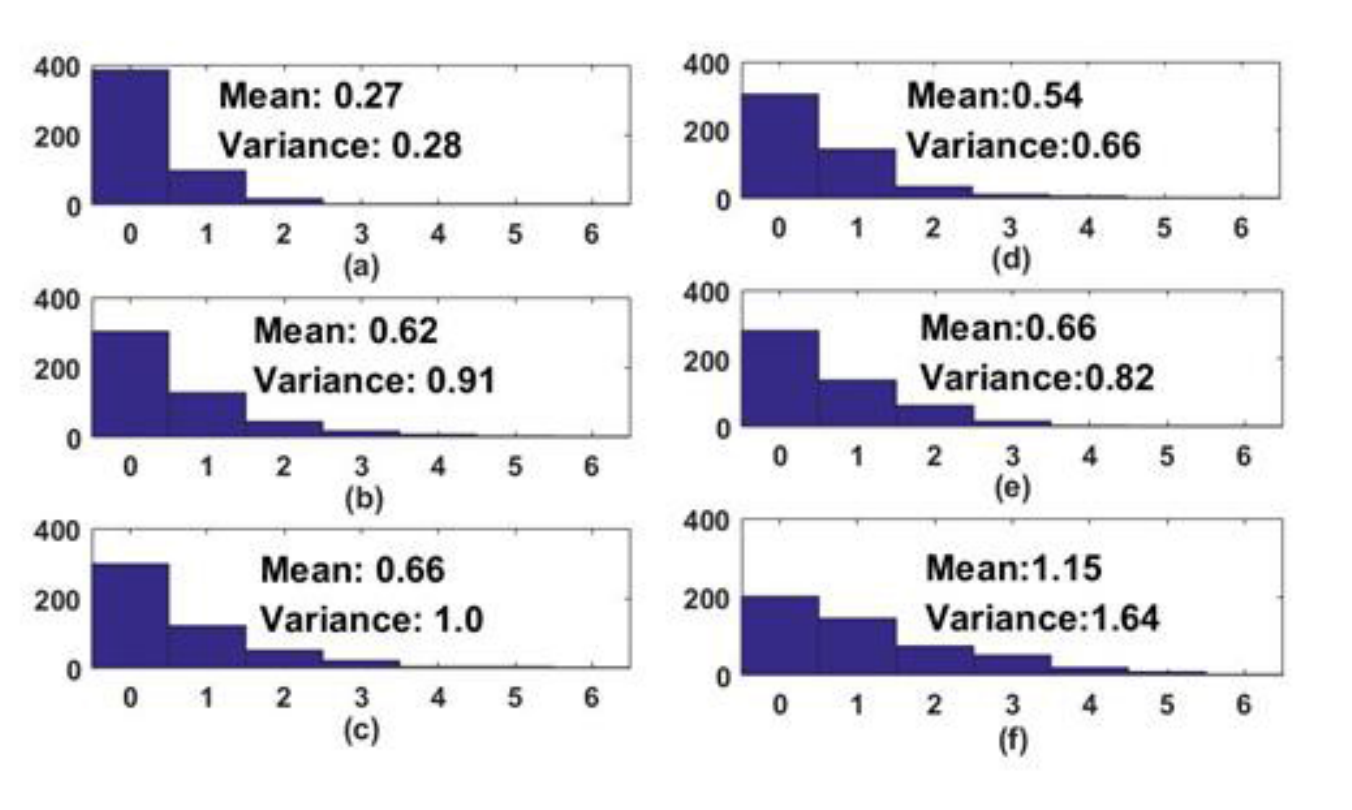,width=1.0\linewidth}

    \caption{The distribution of nanorobots at three different tumor areas T1- (a, d), T2- (b, e), T3- (c, f) using two different MCDP strategies.}
    \label{fig:Fig.13}

\end{center}
\end{figure}

\begin{figure} [!htp]
\begin{center}

\epsfig{file=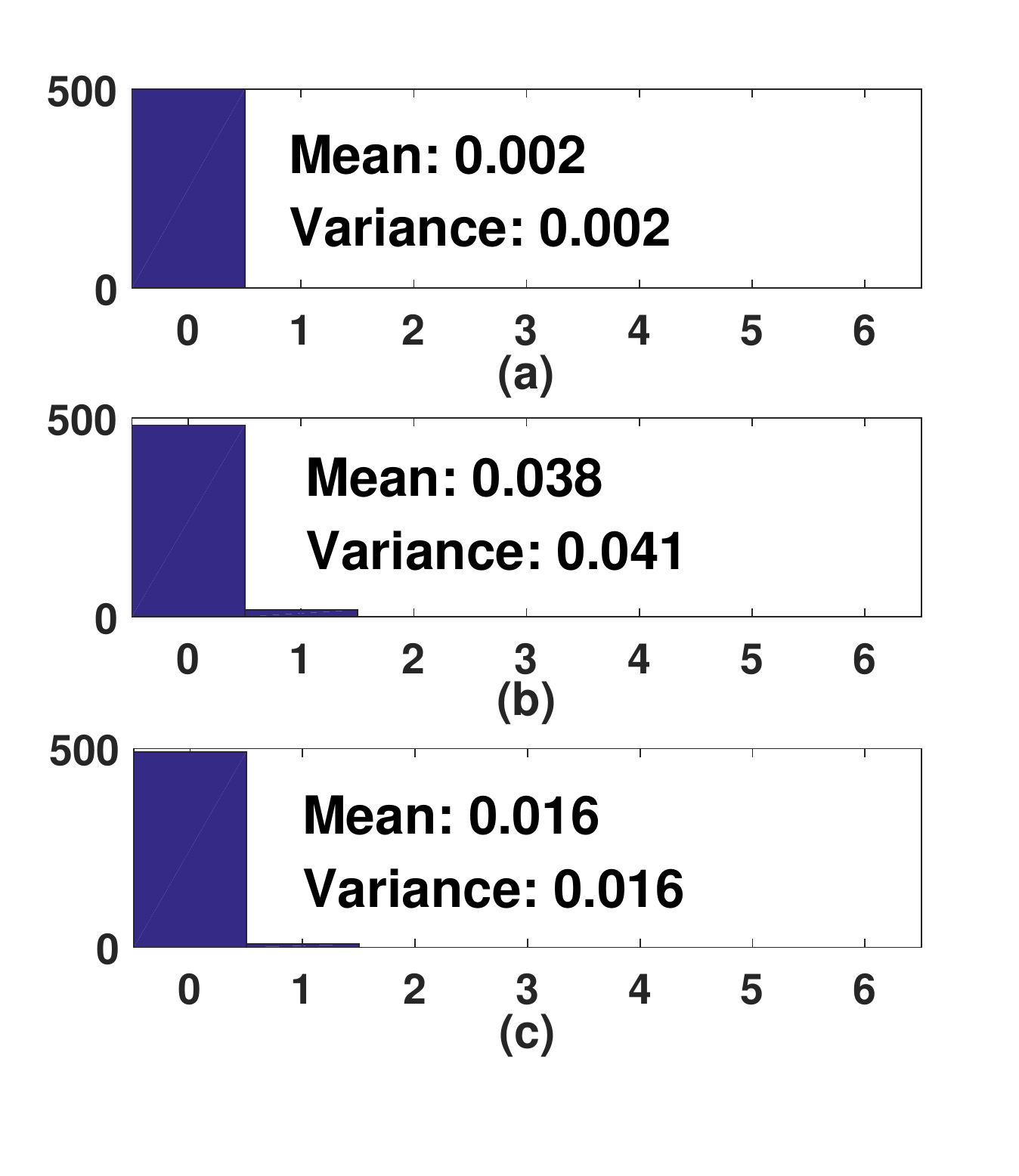,width=0.55\linewidth}

    \caption{The distribution of nanorobots at three different tumor areas T1- (a), T2- (b), T3- (c) using random searching strategy.}
    \label{fig:Fig.14}

\end{center}
\end{figure}
 Fig. 13 presents the histograms of the quantity of nanorobots located at the three different tumors using the two different MCDP strategies. Fig. 13(a), (b) and (c) show the distribution of nanorobots located at the tumor foci T1, T2, and T3 respectively by using the MCDP inspired by the standard NGA.  Fig. 13(d), (e), and (f) show the distribution of nanorobots located at tumor foci T1, T2, and T3 respectively by using the MCDP inspired by the modified NGA. It is evident that for both the two strategies, the numbers of nanorobots in T1, T2, and T3 are increasing, which is in line with the sizes of the tumors T1<T2<T3. For comparison, we also give the distribution of nanorobots located at the tumor areas T1, T2, and T3 respectively by using the random searching strategy, which is shown in Fig. 14. It is interesting to see that the numbers of nanorobots in T1, T2, and T3 are not adaptable to the tumors' area ratio. This is mainly caused by the relative orientations of the three tumors. For example, tumor T1 can be barely detected by any nanorobot under the random searching strategy because it is shadowed by tumor T2 without any line-of-sight to the injected area.

 In summary, we can draw the conclusion that the MCDP performs better than random searching. Furthermore, the MCDP  inspired by the modified NGA performs better than that inspired by the normal NGA.

 \section{Conclusion}
 We have presented the nanorobots-assisted detection of multifocal cancer from a multimodal optimization perspective. Specifically, we have proposed the NGA-inspired MCDP by taking into account realistic $in~vivo$ conditions of nanorobots and characteristics of vascular network around the tumor areas. Based on this work, we have also developed a modified NGA, which can be used to improve the performance of MCDP. Numerical examples have demonstrated the effectiveness of the proposed methodology for the blood flow velocity profile induced by tumor angiogenesis.

 Future work may include improving the performance of the algorithm to accomplish the detection of all the cancer areas with many more nanorobots. It is also important to examine further the impact of nanorobot nonidealities, such as finite lifespan, imprecise steering, and inaccurate tracking.

\ifCLASSOPTIONcaptionsoff
  \newpage
\fi

\bibliographystyle{IEEEtran}
\bibliography{IEEEexample}

\begin{IEEEbiography}{Shaolong Shi}
Biography text here.
\end{IEEEbiography}

% if you will not have a photo at all:
\begin{IEEEbiography}{Yifan Chen}
Biography text here.
\end{IEEEbiography}

% insert where needed to balance the two columns on the last page with
% biographies
%\newpage

\begin{IEEEbiography}{Xin Yao}
Biography text here.
\end{IEEEbiography}

% You can push biographies down or up by placing
% a \vfill before or after them. The appropriate
% use of \vfill depends on what kind of text is
% on the last page and whether or not the columns
% are being equalized.

%\vfill

% Can be used to pull up biographies so that the bottom of the last one
% is flush with the other column.
%\enlargethispage{-5in}

% that's all folks
\end{document}